\documentclass[twocolumn,showpacs,prl,amsfonts,amsmath,amssymb,floatfix,groupedaddress]{revtex4}
\usepackage{color}
\usepackage{hhline}
\usepackage{mathrsfs}
\usepackage{graphicx}
\usepackage{dcolumn}
\usepackage{bm}
\usepackage{multirow}
\usepackage{booktabs}
\usepackage{afterpage}
\usepackage{amsmath}
\usepackage{ulem}

\begin{document}

\title{Possible ``Magn\'eli" phases and self-alloying in the superconducting sulfur hydride }

\author{Ryosuke Akashi$^{1}$}
\thanks{akashi@cms.phys.s.u-tokyo.ac.jp}
\author{Wataru Sano$^{2,3}$}
\author{Ryotaro Arita$^{3,4}$}
\author{Shinji Tsuneyuki$^{1,5}$}
\affiliation{$^1$Department of Physics, The University of Tokyo, Hongo, Bunkyo-ku, Tokyo 113-0033, Japan}
\affiliation{$^2$Department of Applied Physics, The University of Tokyo, Hongo, Bunkyo-ku, Tokyo 113-8656, Japan}
\affiliation{$^3$RIKEN Center for Emergent Matter Science, Wako, Saitama 351-0198, Japan}
\affiliation{$^4$JST ERATO Isobe Degenerate $\pi$-Integration Project, Advanced Institute for Materials Research (AIMR), Tohoku University, Sendai, Miyagi 980-8577, Japan}
\affiliation{$^5$Institute of Solid State Physics, The University of Tokyo, Kashiwa, Chiba 277-8581, Japan}

\date{\today}
\begin{abstract}
We theoretically give an infinite number of metastable crystal structures for the superconducting sulfur hydride H$_{x}$S under pressure. Previously predicted crystalline phases of H$_{2}$S and H$_{3}$S have been thought to have important roles for the experimentally observed low and high $T_{\rm c}$, respectively. The newly found structures are long-period modulated crystals where slab-like H$_{2}$S and H$_{3}$S regions intergrow in a microscopic scale. The extremely small formation enthalpy for the H$_{2}$S--H$_{3}$S boundary indicated with the first-principles calculations suggests possible alloying of these phases through formation of local H$_{3}$S regions. The modulated structures and gradual alloying transformations between them explain peculiar pressure dependence of $T_{\rm c}$ in sulfur hydride observed experimentally, as well as could they prevail in the experimental samples under various compression schemes. 
\end{abstract}

\maketitle
Sulfur hydride has recently been found to become superconductor at extremely high pressure around 200~K~\cite{Eremets, Eremets-arxiv}. This is the first achievement of superconducting transition temperature ($T_{\rm c}$) exceeding the nitrogen boiling point among the conventional phonon mediated superconductors~\cite{BCS}, as well as it has broken the long-standing record of 160~K in mercury-cuprate~\cite{Chu-cuprate-press-Nat,Gao-cuprate-press-PRB,Monteverde-cuprate-press-EPL,Takeshita-cuprate-press-JPSJ}. 

\begin{figure}[b!]
 \begin{center}
 \includegraphics[scale=0.7]{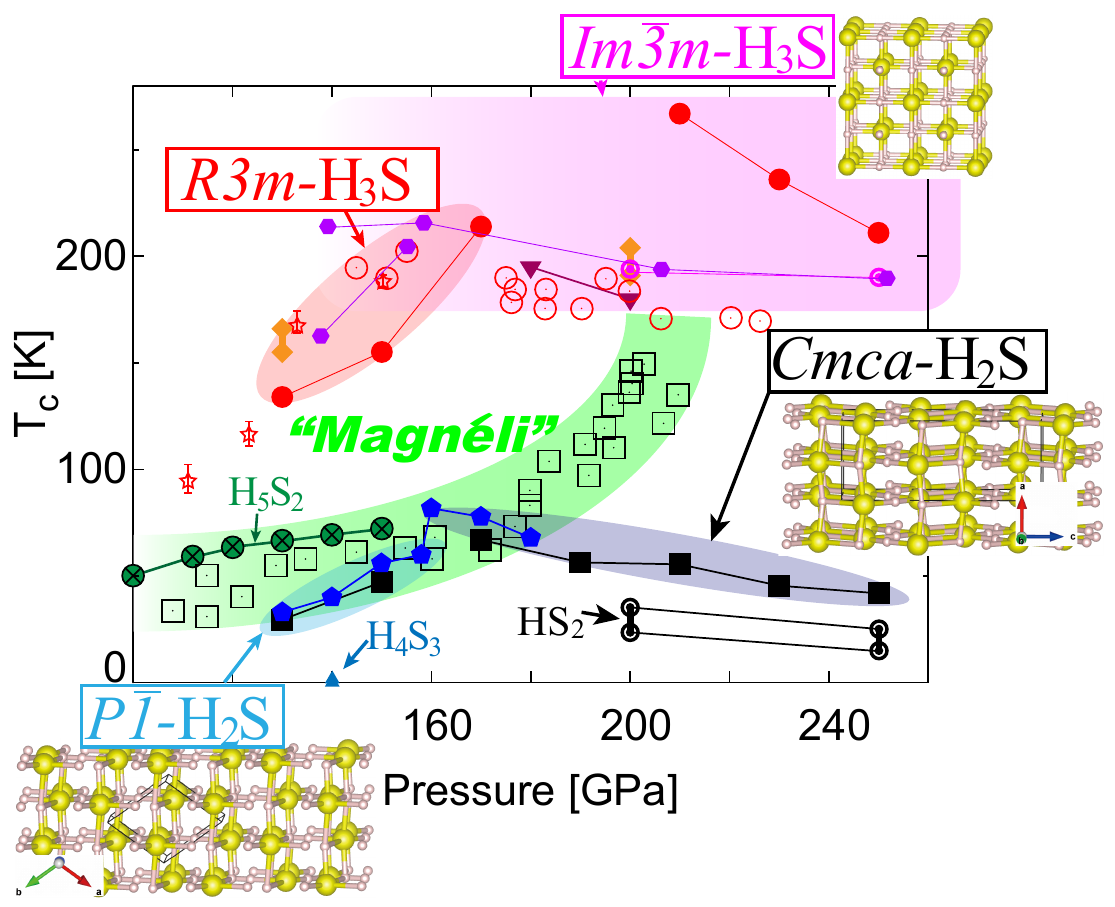}
  \caption{Experimentally observed $T_{\rm c}$ (open circle, square~\cite{Eremets} and star with error bar~\cite{Shimizu-struct-arxiv}) compared with the first-principles calculations (solid symbols). The latter data are taken from Ref.~\onlinecite{Li-Ma-H2S-struct} (pentagon), Ref.~\onlinecite{Duan-H3S-struct} (rhombus), Ref.~\onlinecite{Flores-Livas-SCDFT} (inverted triangle), Ref.~\onlinecite{Akashi-HxS} (circle and square), Ref.~\onlinecite{Errea-anharm-PRL} (double circle), Ref.~\onlinecite{Li-Errea-various-struct-arxiv} (triangle), Ref.~\onlinecite{Errea-symmetrize-Nat} (hexagon) and Ref.~\onlinecite{Ishikawa-H5S2} (circle with x-mark). The data points in the shaded areas have been obtained from the corresponding predicted crystal structures~\cite{Li-Ma-H2S-struct,Duan-H3S-struct}, respectively. Note that $R3m$-H$_{3}$S is a trigonally distorted variant of $Im\bar{3}m$-H$_{3}$S. As discussed later, we attribute the pressure dependence of $T_{\rm c}$ (open square) to the ``Magn\'{e}li" phases. The values from Ref.~\onlinecite{Akashi-HxS} (Refs.~\onlinecite{Errea-anharm-PRL,Errea-symmetrize-Nat}) includes the plasmon effect (phonon anharmonic effect). The vertical bars connecting the rhombi indicate the $T_{\rm c}$ variation with the empirical Coulomb parameter $\mu^{\ast}$~\cite{Duan-H3S-struct,Errea-anharm-PRL}.}
  \label{fig:crystals}
 \end{center}
\end{figure}

A remarkable feature observed in this superconducting phenomenon is pressure and annealing-scheme dependences of $T_{\rm c}$~\cite{Eremets, Eremets-arxiv,Shimizu-struct-arxiv}. (i) When the pressure ($>$$\sim$100~GPa) is applied to the H$_{2}$S sample at room temperatures and afterwards cooled down, the observed $T_{\rm c}$s amount to over 150 K, with 203K being the maximum value (open circle in Fig.~\ref{fig:crystals}). (ii) By pressurizing at temperature around 100 K, on the other hand, the observed $T_{\rm c}$ remains low and next rapidly increases (open square in Fig.~\ref{fig:crystals}). Although this behavior suggests a variety of structural phases and their peculiar properties, efficient experimental observations have been obstructed by the extremely high pressure. Instead, first-principles calculations have provided insights for the superconducting phases. It is now established that some of the observed values of $T_{\rm c}$ with the low- and high-$T$ schemes are well reproduced~\cite{Li-Ma-H2S-struct, Duan-H3S-struct,Mazin-enthalpy,Flores-Livas-SCDFT,Akashi-HxS,Errea-anharm-PRL,Errea-symmetrize-Nat,Ishikawa-H5S2,Sano-HxS} with crystal structures predicted for compositions of H$_{2}$S~\cite{Li-Ma-H2S-struct}, as well as H$_{5}$S$_{2}$~\cite{Ishikawa-H5S2} and H$_{3}$S~\cite{Duan-H3S-struct}~(See Fig.~\ref{fig:crystals}). However, consistent understanding on the observed behavior still remains unprecedented. In particular, no clear explanation on the rapid increase of $T_{\rm c}$ with the low-$T$ scheme has been given, despite accumulated first-principles proposals of candidate structures with various composition of H$_{x}$S~\cite{Mazin-enthalpy,Duan-various-struct,Errea-anharm-PRL,Li-Errea-various-struct-arxiv}.

The current theoretical attempts have focused on the possible understanding with a minimal number of distinct structural phases. Rather, we provide a different view: Not only distinct phases but also their mixture have vital roles. Specifically, we find an infinite number of metastable crystal structures having compositions H$_{x}$S$_{1-x}$ with 2/3$<$$x$$<$3/4, which have not been reported in the first-principles structure-search studies. They can be understood as long-period modulated crystals formed by stacking the H$_{2}$S and H$_{3}$S slab-like structures, which are reminiscent of the Magn\'{e}li phases in transition-metal oxides~\cite{Magneli}. All these structures are thermodynamically as stable as the complete separation to H$_{2}$S and H$_{3}$S phases. This suggests that the microscopic {\it intergrowth} of the H$_{2}$S and H$_{3}$S regions requires little activation enthalpy and therefore occurs ubiquitously in the experimental situation, forming, as it were, slab-alloy phase. We also show that the experimentally observed $T_{\rm c}$-pressure curve is reproduced from these alloy-like phases.

\begin{figure}[t!]
 \begin{center}
  \includegraphics[scale=0.27]{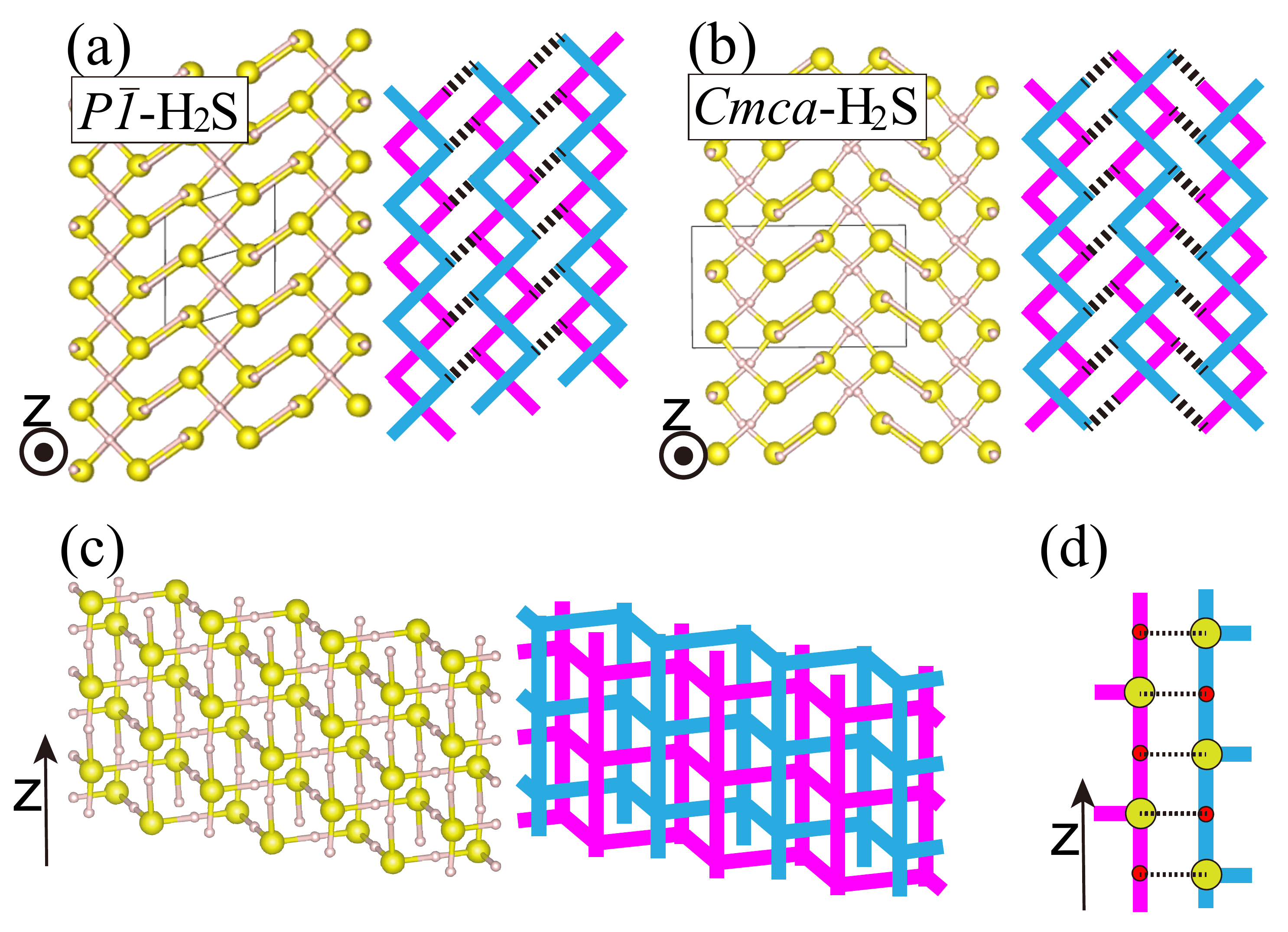}
  \caption{(a) (1 1 0) view of $P\bar{1}$-H$_2$S and (b) (1 0 0) view of $Cmca$-H$_{2}$S, accompanied by their schematic pictures. The unit cells are depicted in thin lines. In the schematic pictures, the bond depicted by panel (d) is indicated by thick dashed lines. (c) The fundamental unit structure, ``interlaced square nets" and its schematic picture. In the latter, sulfur and hydrogen atoms are located at the corners and the centers of the edges, respectively. (d) Bonding structure in panels (a) and (b), where sulfur and hydrogen atoms are depicted in yellow and red, respectively. } 
  \label{fig:relations}
 \end{center}
\end{figure}

\begin{figure}[b!]
 \begin{center}
  \includegraphics[scale=0.35]{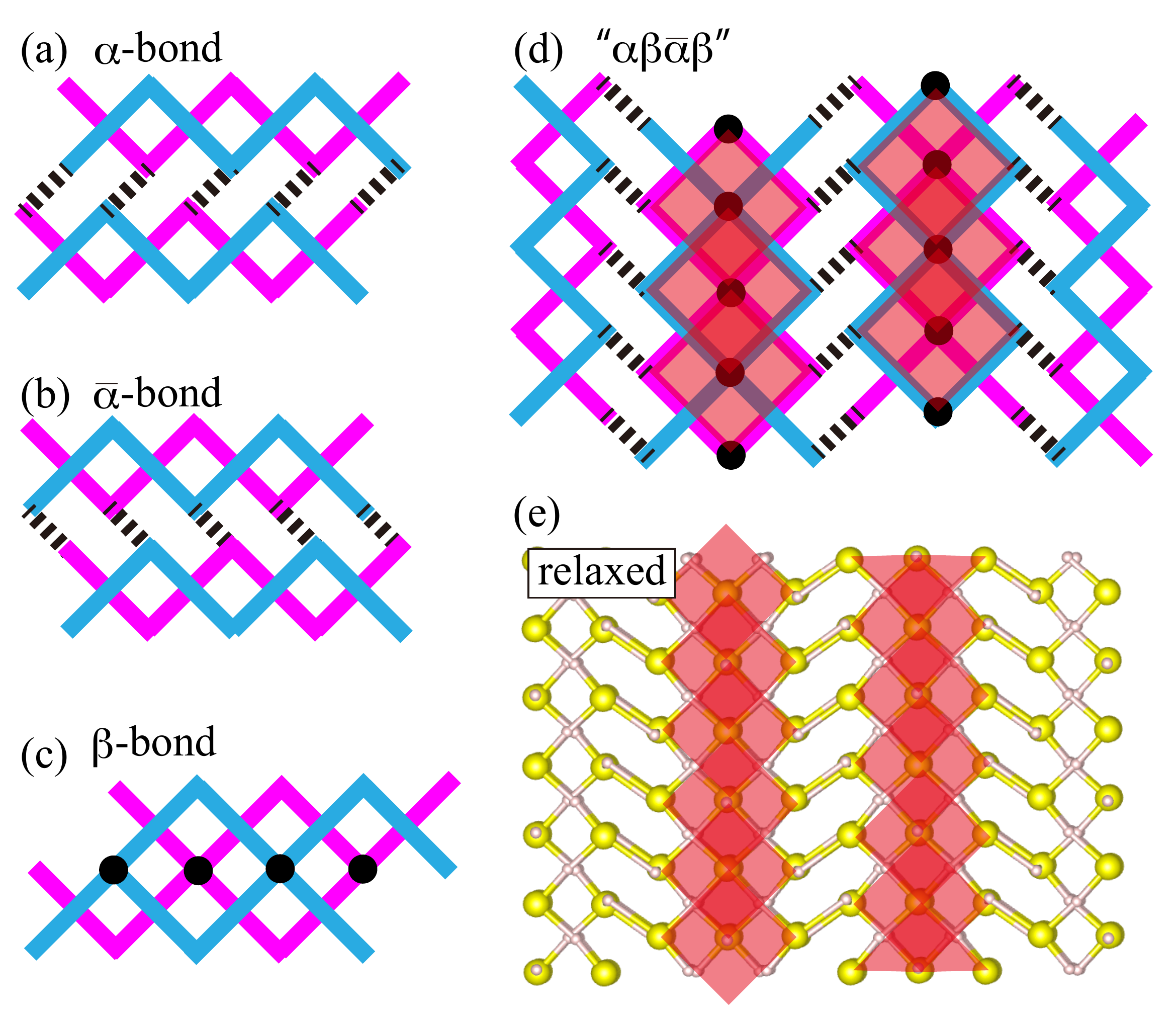}
  \caption{(a)--(c) Definitions of the inter-unit bonding. (d) Long-period modulated structure generated from permutation ``$\alpha \beta \bar{\alpha} \beta$" and (e) its relaxed counterpart (see the main text). The $Im\bar{3}m$-H$_3$S regions are shaded in red.}
  \label{fig:unit-bond}
 \end{center}
\end{figure}

We begin with a discussion of the hidden similarity between theoretically predicted relevant structures for H$_2$S and H$_3$S. $P\bar{1}$-H$_{2}$S and $Cmca$-H$_{2}$S (Fig.~\ref{fig:crystals}) are known to give low $T_{\rm c}$ values~\cite{Li-Ma-H2S-struct, Akashi-HxS,Sano-HxS}, which agree relatively well with the experimental values with the low-$T$ annealing for the low-pressure regime. $Im\bar{3}m$-H$_{3}$S  (Fig.~\ref{fig:crystals}) has been thought to give the $T_{\rm c}$ with the high-$T$ annealing cases~\cite{Errea-anharm-PRL, Errea-symmetrize-Nat, Pickett-atomic, Nicol-Carbotte,Boeri-H3Ch,Komelj-hybrid, Bianconi-NovSM,Bianconi-review,Bianconi-Scirep,Sano-HxS,Mazin-enthalpy,Pickett-Wannier,Ortenzi-local,Gorkov-Scirep,Durajski-Pietronero-H3S}, and its presence is recently confirmed with the X-ray diffraction measurements~\cite{Shimizu-struct-arxiv, comment-distortion, Li-Errea-various-struct-arxiv,Goncharov-struct-arxiv}. Seen from the (1 1 0) direction [(1 0 0) direction] in the setup of Ref.~\onlinecite{Li-Ma-H2S-struct}, one can notice that $P\bar{1}$-  and $Cmca$-H$_2$S can be decomposed into the common unit structures [Figs.~\ref{fig:relations}(a)(b)]: Two H$_2$S square nets interlaced with each other [Fig.~\ref{fig:relations}(c)]. The units of this shape are bound with each other so that sulfur atoms adopt local FCC-like stacking [Fig.~\ref{fig:relations}(d)]~\cite{comment-bonding}. The $P\bar{1}$ and $Cmca$ structures can now be distinguished by the configuration of the inter-unit bonding. The uniform (alternating) bonding orientation yields the $P\bar{1}$ ($Cmca$) structure. An important thing is that the $Im\bar{3}m$-H$_3$S structure can also be formed with the present unit by binding them so that the vertical edges are shared [Fig.~\ref{fig:unit-bond}(c)].

The unit-based perspective for H$_2$S and H$_3$S is indeed helpful for us to construct a group of metastable structures. In a wide variety of materials such as metal oxides, multiple crystalline phases with slightly different stoichiometries emerge depending on the bonding between the unit complexes~\cite{Bragg-book,Eyring-Tai-book}. One of such examples is the Magn\'{e}li phases in molybdenum-~\cite{Magneli,Bursill-PRSoc-TEM, Wang-TEM-hyreso}, tungsten-~\cite{Magneli, Tilley-TEM}, titanium-~\cite{Andersson-Magneli,Terasaki-TEM,Bursill-Hyde-review, Harada-Thermo}, and vanadium-oxides~\cite{Andersson-Jahnberg-VOx,Hirotsu-VOx-highreso,Canfield-phasediagram, Schwingenschlogl-AnnPhys}, where two-dimensional defects (or {\it crystallographic shear}~\cite{Wadsley-review}) of edge- or face-sharing bonding between metal-oxide complexes are periodically formed. On the analogy to them, here we consider possible crystalline phases formed by the structural H$_2$S unit. 

We first define the three types of the inter-unit bonding (Fig~\ref{fig:unit-bond} (a)--(c)): bonding $\alpha$ and $\bar{\alpha}$ correspond to the two bonding orientations seen in $Cmca$-H$_2$S, respectively, whereas bonding $\beta$ corresponds to the bond sharing the edge. With this definition, by arranging the units side by side and next assigning any type of bonding for every neighboring pair of units, we can generate a crystal structure formed by the units. In this sense, every circular permutation composed of desired numbers of $\alpha$, $\bar{\alpha}$ and $\beta$ yields a corresponding long-period modulated crystal structure. Any structures generated in this way have intermediate composition H$_x$S$_{1-x}$ ($2/3\leq x\leq 3/4$), whose unit cells are composed of H$_{4N-k}$S$_{2N-k}$ with $N$ and $k$ being the period of the permutation and the number of $\beta$, respectively. The ``$\alpha \beta \bar{\alpha} \beta$" structure is exemplified in Fig.~\ref{fig:unit-bond}(d), whose unit-cell formula is H$_{14}$S$_{6}$. As the number of $\beta$ increases, H$_3$S-like structural regions grow [Fig.~\ref{fig:unit-bond} (d)(e)]. Note that $P\bar{1}$-H$_2$S, $Cmca$-H$_2$S, and $Im\bar{3}m$-H$_3$S are generated by permutations ``$\alpha$" (also ``$\bar{\alpha}$"; see Supplemental Materials~\cite{comment-suppl}), ``$\alpha\bar{\alpha}$", and ``$\beta$", respectively. 

\begin{figure}[t!]
 \begin{center}
  \includegraphics[scale=0.65]{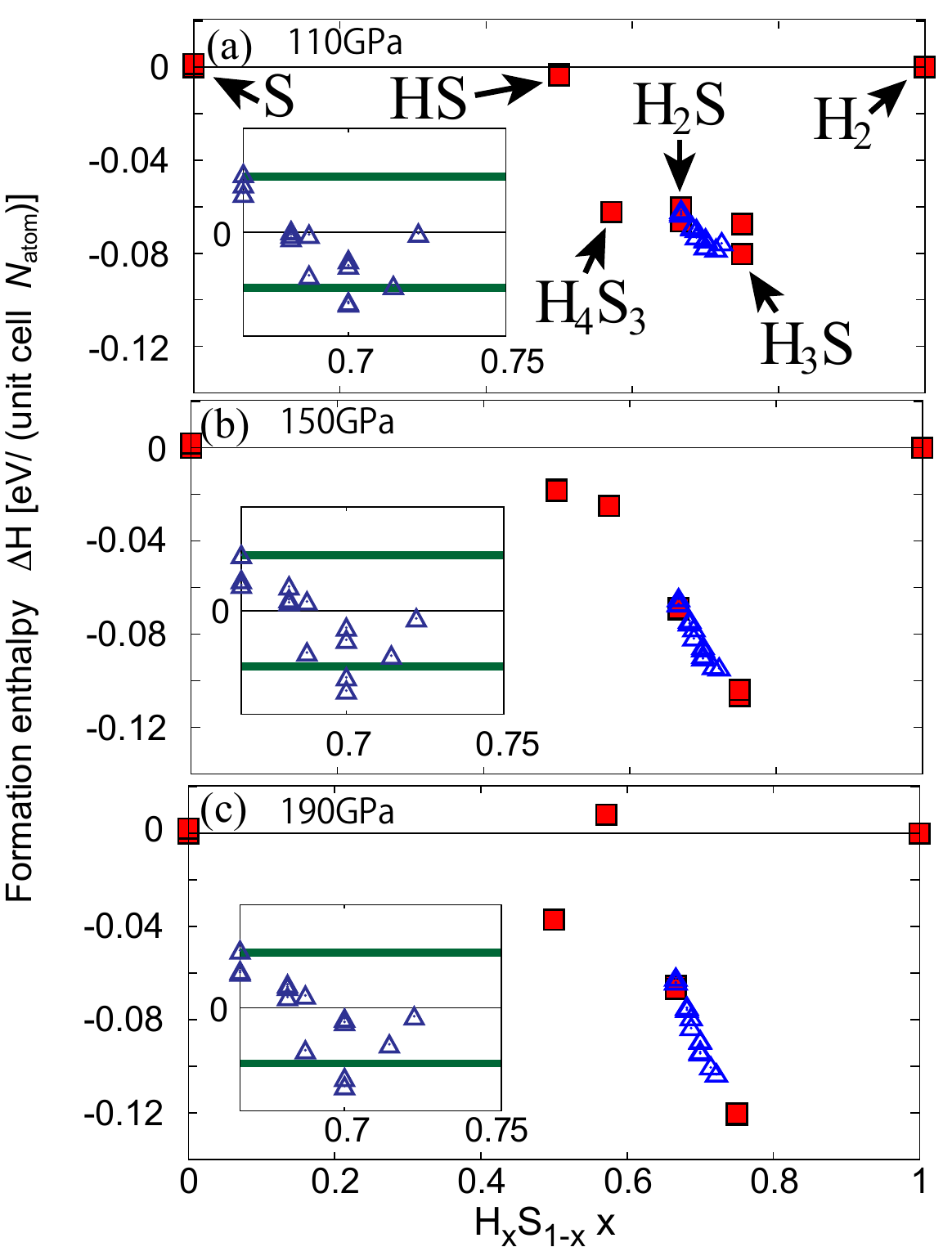}
  \caption{(a)--(c) Formation enthalpy from the first principles at various pressures, where $N_{\rm atom}$ denotes the number of atoms in the unit cell. The structures corresponding to the points are summarized in Supplemental Materials~\cite{comment-suppl}. The inset panels focus the region between $x=2/3$ (H$_2$S) and $x=3/4$ (H$_3$S), where the formation enthalpies are measured with respect to the values for the phase separation of H$_2$S and H$_3$S. Bold lines indicate $\pm$50 kelvin.}
  \label{fig:enthalpy}
 \end{center}
\end{figure}

Using the structures thus generated from the permutations of period 2, 3, and 4 as inputs, we carried out the first-principles structure optimization for various pressures. Our calculations were done using the first-principles code package {\sc QUANTUM ESPRESSO}~\cite{QE} with the generalized-gradient approximation for the exchange-correlation potential~\cite{GGAPBE}. The unit-cell compositions of the resulting structures are H$_7$S$_3$, H$_{10}$S$_{4}$, H$_{11}$S$_{5}$,H$_{12}$S$_{6}$, H$_{13}$S$_{5}$, H$_{14}$S$_{6}$, H$_{15}$S$_{7}$, and H$_{16}$S$_{8}$, respectively. Detailed conditions of the calculations are summarized in Supplemental Materials~\cite{comment-suppl}. We have found that the formation enthalpy $\Delta H$(H$_{x}$S$_{1-x}$) for all the resulting structures satisfy $\Delta H(\rm H_{2}S) \gtrsim \Delta H(\rm H_{\it x}S_{1-{\it x}}) \gtrsim \Delta H(\rm H_{3}S)$ (Fig.~\ref{fig:enthalpy}). We have also confirmed that all these structures retain the bonding characteristics in their initial structures [Fig.~\ref{fig:unit-bond}(e)]. Namely, every different permutation gives different optimum metastable structures where well-defined bcc-H$_3$S slab regions emerge. Although we do not examine the $N$$\geq$$5$ cases here, an infinite number of metastable structures are expected to be obtained in this way.

Remarkably, the values of formation enthalpy relative to the complete decomposition into H$_2$S and H$_3$S are wholly within 50~K per atom [insets of Fig.~\ref{fig:enthalpy}(a)--(d)]. This dependence is not an indication of trivial phase separation into H$_2$S and H$_3$S because the two regions {\it intergrow} in a microscopic scale in the respective structures (see Fig.~\ref{fig:unit-bond} (e)). A more appropriate interpretation is that the formation enthalpy for the H$_2$S--H$_3$S boundary is extremely small in these structures. 

Let us argue the relevance of the newly found intermediate phases in the experimental situation.  We here have to bear in mind that the intermediate structures are {\it within} the convex hull of the formation enthalpy in the present pressure regime, apart from the narrow range around 110 GPa (Fig. \ref{fig:enthalpy} and Ref.~\onlinecite{Li-Errea-various-struct-arxiv}). This means that in the compressed H$_2$S system, these phases do not grow as the decomposition residues of the emergence of H$_3$S phase, if annealed with enough time and heat. However, the decomposition paths otherwise depend on the enthalpy barrier from the pristine H$_2$S structures. The transition between two metastable structures with slightly different $x$ occurs through transformations of a small fraction of $\alpha~(\bar{\alpha})$-type inter-unit bonds to $\beta$. This sporadic character of the bonding transformation and the small formation enthalpy for the H$_2$S--H$_3$S boundary revealed above suggest that such transition requires little activation enthalpy. The intermediate crystalline phases should thus be observable if H$_{x}$S systems exhibit either of the $P\bar{1}$- or $Cmca$-H$_{2}$S region and next further compressed at low temperature. 

The possible emergence of the intermediate Magn\'{e}li-type phases draws a consistent explanation on the puzzling pressure dependence of $T_{\rm c}$ observed experimentally. In the present ``Magn\'{e}li" phases, $T_{\rm c}$ is expected to increase as the H$_{3}$S regions grow. Indeed, by selecting plausible phases and calculating their superconducting $T_{\rm c}$s from the first principles~\cite{Migdal-Eliashberg,Baroni,opt-tetra,SCDFTI, SCDFTII,Akashi-Z-asym,Akashi-plasmon-PRL,Akashi-plasmon-JPSJ,Akashi-HxS,QE,comment-suppl}, we successfully reproduced the experimentally observed $T_{\rm c}$ behavior~(Fig.\ref{fig:crystals}): 41~K (130GPa), 80~K (180GPa), 107~K (190GPa) and 121~K (200GPa). Here, the increase of $T_{\rm c}$ is due to enhancement of electron-phonon coupling~(see Supplemental Materials~\cite{comment-suppl}). The succession of the ``Magn\'{e}li" phases with increasing fraction of the H$_{3}$S region is hence the probable origin of the pressure dependence of $T_{\rm c}$ observed with the low-$T$ compression scheme, filling the unsolved gap between the theory and experiments. Although we have selected the ``$\alpha \beta$", ``$\alpha \beta \beta$",, ``$\alpha \beta \beta \beta$", and ``$\alpha$($\beta$$\times$9)" phases for 130, 180, 190 and 200~GPa, respectively, we do not conclude these specific phases were emerging in the experimental situation in Ref.~\onlinecite{Eremets} but were more various Magn\'{e}li phases. For example, there has been a recent first-principles calculation that H$_{5}$S$_{2}$, which is actually equivalent to the ``$\alpha \beta \beta$" phase, also yields a good agreement for $P$$<$$150$~GPa (circle with x-mark in Fig.~\ref{fig:crystals}; Ref.~\onlinecite{Ishikawa-H5S2}). 
\begin{figure}[t!]
 \begin{center}
  \includegraphics[scale=0.35]{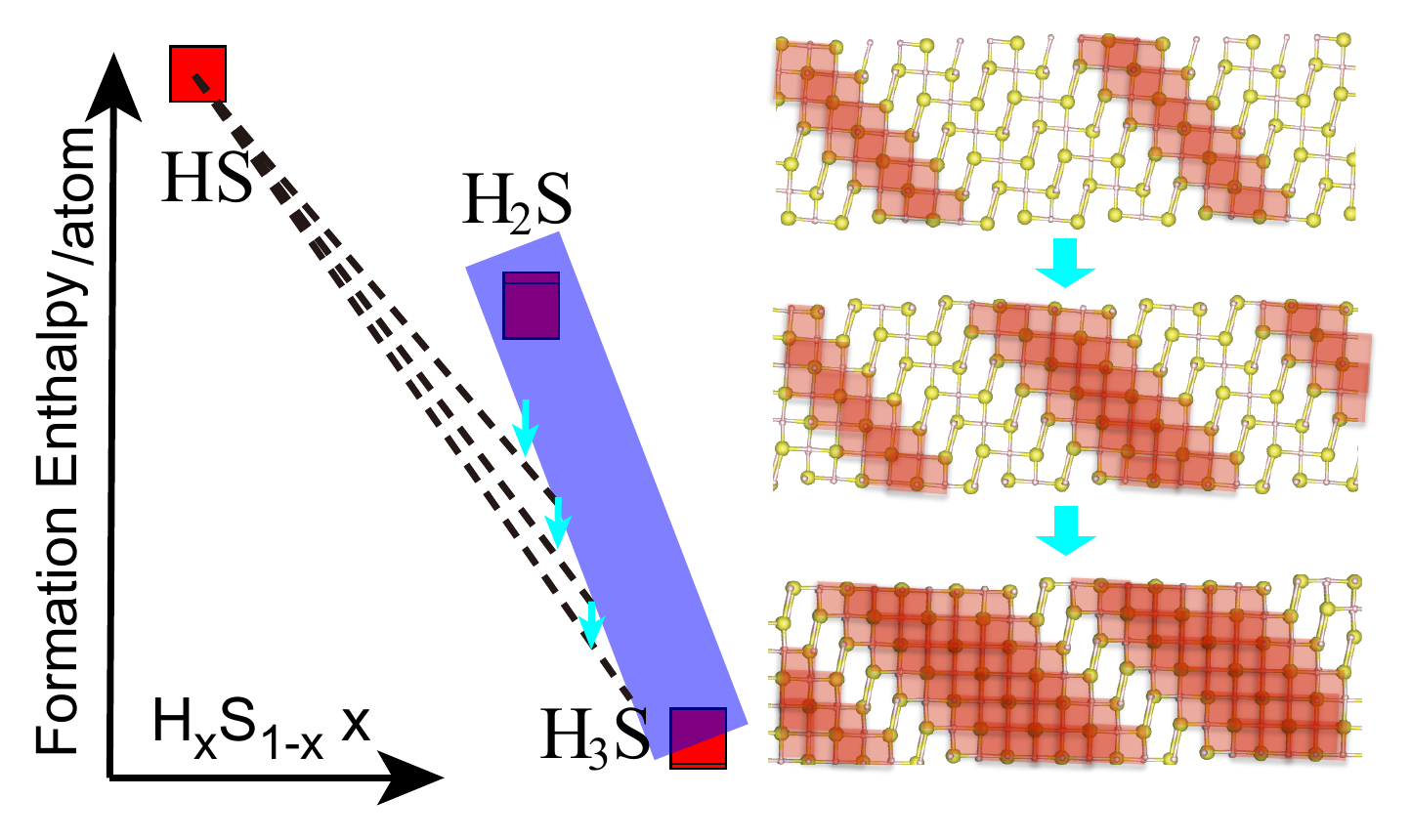}
  \caption{Schematic picture of a possible transformation path ``$\alpha$ ($\bar{\alpha}$) $\rightarrow$ $\beta$+HS". The shaded area in the left plot represents the multiple intermediate structures, whereas that in the right indicate the H$_{3}$S regions.}
  \label{fig:transform-scheme}
 \end{center}
\end{figure}

\begin{figure}[t!]
 \begin{center}
  \includegraphics[scale=0.75]{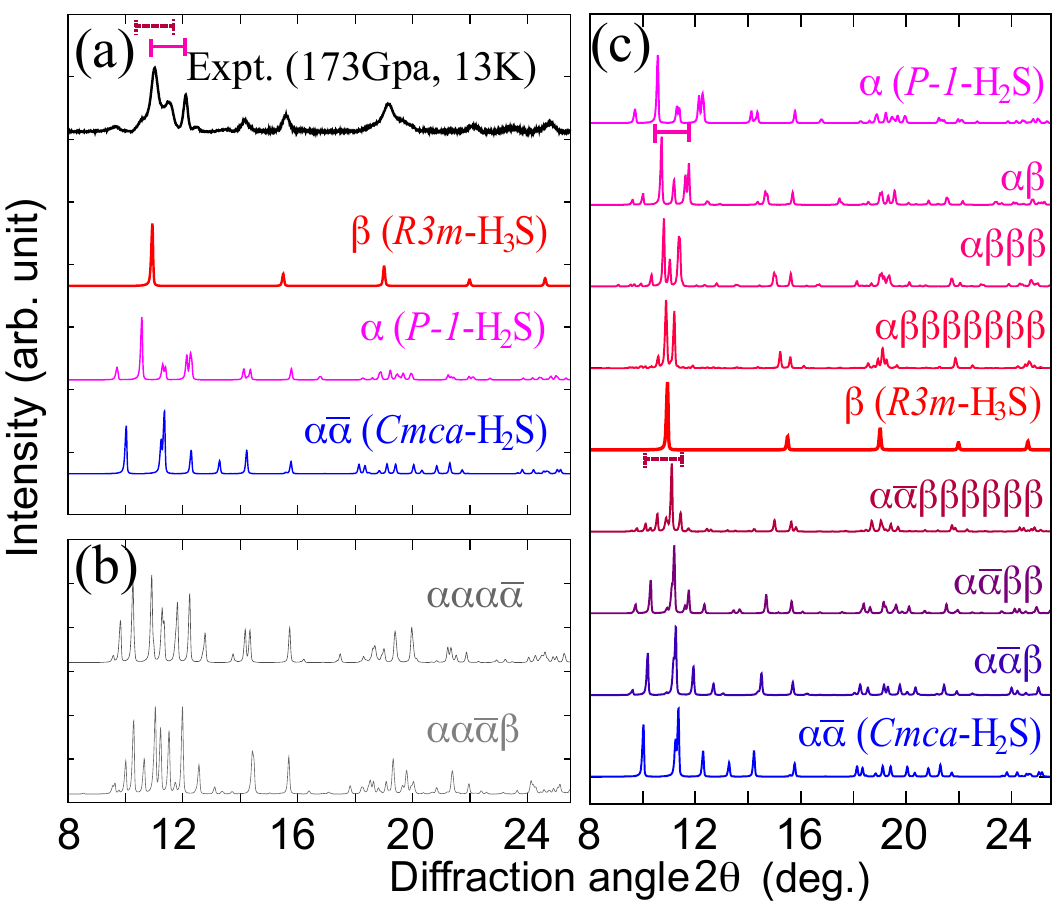}
  \caption{X-ray diffraction patterns. The simulations were done for the structures optimized at 150 GPa. (a) (top) Experimentally observed diffraction pattern for D$_{2}$S taken from Ref.~\onlinecite{Shimizu-struct-arxiv}, where the horizontal bars indicate regions where we can find similarity to simulated patterns for some Magn\'{e}li phases [see panel (c)]. (bottom) The simulated patterns for the reference structures. (b) Representative simulated patterns exhibiting multiple diffraction peaks. (c) Patterns obtained from a selected group of structures which are ordered so that the fraction of $\beta$-bond gradually varies.}
  \label{fig:Xray}
 \end{center}
\end{figure}

We can also find a plausible reaction path consistent with the experiment. The calculated formation enthalpies for HS and the ``Magn\'{e}li" phases decrease with compression compared with those for H$_{2}$S~[see Fig.~\ref{fig:enthalpy} (a)--(c)]. This pressure-induced phenomenon would stimulate a local reactions forming the H$_{3}$S slab: $\alpha$ ($\bar{\alpha}$)$\rightarrow$$\beta$+HS (Fig.~\ref{fig:transform-scheme}; see also Supplemental Materials~\cite{comment-reaction}). This supports the present scenario; the increase of the H$_{3}$S region by compression. 

The multiplicity of the metastable phases deduces an interesting speculation: Arbitrary values of $T_{\rm c}$ between those of the low- and high-$T_{\rm c}$ phases in Ref.~\onlinecite{Eremets} are observable. The local H$_{3}$S-slab formation can also be stimulated by annealing. By carefully controlling the annealing conditions, one would observe various values and their temporal evolution even at fixed pressures.

Here we note a possible characteristic effect in the present H$_{x}$S Magn\'{e}li phases for future studies. In the titanium-oxide Magn\'{e}li phase, it is known that the local electron-lattice coupling is enhanced by the two-dimensional defects to form bipolarons (or charge order)~\cite{Lakkis-bipolaron}. Hydrogen atoms are in principle subject to this instability because of their multivalent character. Although we have not found traces of such polaronic phases~(see Supplemental Materials~\cite{comment-suppl}), in view of the strong electron-phonon coupling, proximity effects of such phases may affect the transport and superconducting properties~\cite{Chakraverty-bipolaronSC,Alexandrov-polaronSC,Alexandrov-bipolaronSC,Alexandrov-review}. 

To facilitate experimental exploration of the H$_x$S Magn\'{e}li phases, we computed theoretical X-ray diffraction patterns at 150GPa using RIETAN package (Ref.~\onlinecite{RIETAN}, see Supplemental Materials\cite{comment-suppl}). We focus on two distinct groups of structures. When the fraction of the $\beta$ bond is small and both $\alpha$ and $\bar{\alpha}$ bonds are present, multiple minor peaks are generally seen around the diffraction angle of the dominant peak in $R3m$-H$_{3}$S [(110) peak; $2$$\theta$$\simeq$$11^{\circ}$ in Fig.~\ref{fig:Xray}(a)], as exemplified in Fig.~\ref{fig:Xray}(b). This behavior is due to subtle distortion of the sulfur lattice and apparently difficult to understand with a simple combination of the patterns for the pristine $R3m$-H$_{3}$S, $P\bar{1}$-H$_{2}$S and $Cmca$-H$_{2}$S phases. Experimental ``noise" may actually be contributed to by the diffraction peaks from those phases. 
We also show the simulated patterns when the $\beta$ bonds are gradually introduced in either $P\bar{1}$- or $Cmca$-H$_{2}$S phase in Fig.~\ref{fig:Xray}(c). We observe gradual evolutions between the end points. Remarkably, the intensity of the dominant peak is well retained while peripheral features such as sub-peaks, peak shift and tail are found. In the previous diffraction experiments~\cite{Shimizu-struct-arxiv,Li-Errea-various-struct-arxiv,Goncharov-struct-arxiv}, presence of the H$_{3}$S phase has been indicated from the (110) peak, though it was accompanied by subtle sub-peaks depending on the experimental protocol. The Magn\'{e}li phases of this group may provide a unified understanding on such structures. To demonstrate this, we plotted an experimental pattern quoted from Ref.~\onlinecite{Shimizu-struct-arxiv} in Fig.~\ref{fig:Xray}(a). The sub-peaks and tails around $10^{\circ}$$ \lesssim$$ 2\theta$$ \lesssim $$12^{\circ}$ well resembles to, for example, those for $\alpha\beta$ and $\alpha\bar{\alpha}\beta\beta\beta\beta\beta\beta$, as indicated in panels (a) and (c). 

We additionally assert that not only crystalline Magn\'{e}li phases but also their nanoscale stripes can emerge in an ordered and random fashion, similarly to the cases of the microsyntactic intergrowth~\cite{Hirotsu-VOx-highreso} observed in silicon carbide~\cite{Sato-SiC} and metallic oxides including the Magn\'{e}li materials~\cite{Bevan-alumina, Hirotsu-ferrite, Hirotsu-vanadate, Hirotsu-vanadate2,Hirotsu-VOx-highreso}. Such structures will appear in reality an alloy-like phase formed by the slabs of the metallic H$_2$S and H$_3$S. This alloy phase is a compound analog of the classic superconducting alloys formed by elemental metals~\cite{Matthias-review}, but quite different from them. The ingredients of the former alloy---H$_{2}$S and H$_{3}$S slabs---can develop in the common H$_{2}$S crystalline phases and therefore, even in the pristine sample, their alloying occurs in the self-contained manner through the structural and stoichiometric transformations. This alloying obviously smears the experimental diffraction pattern, which could appear to be amorphous-like behavior. 

In summary, we have provided an interpretation based on the common structural unit for the known crystal structures of the low- and high-$T_{\rm c}$ H$_2$S and H$_3$S and thereby found metastable Magn\'{e}li-type structures formed by their layered microsyntactic intergrowth. The experimentally observed pressure dependence of $T_{\rm c}$ is reasonably explained with pressure-induced gradual transformations through such alloy phases. The present finding gives a new insight into the high-$T_{\rm c}$ superconducting phenomena in sulfur hydride that the microscopic mixture of the phases can be ubiquitous.

{\it -Acknowledgment.}
This work was supported by MEXT Element Strategy Initiative to Form Core Research Center in Japan and JSPS KAKENHI Grant Numbers 15K20940 (R. Ak) and 15H03696 (R. Ar) from Japan Society for the Promotion of Science (JSPS). We thank Yinwei Li and Mari Einaga for sharing structure data for H$_4$S$_3$ in Ref.~\onlinecite{Li-Errea-various-struct-arxiv} and X-ray diffraction data in Ref.~\onlinecite{Shimizu-struct-arxiv}, respectively. We also thank Mikhail Eremets, Yoshihiro Iwasa, Hui-hai Zhao, and Yuta Tanaka for enlightening comments. The crystal structures were depicted using VESTA~\cite{VESTA}. The superconducting properties were calculated at the Supercomputer Center at the Institute for Solid State Physics in the University of Tokyo.

\clearpage
\section{SUPPLEMENTAL MATERIALS}

\begin{figure}[b!]
 \begin{center}
  \includegraphics[scale=0.25]{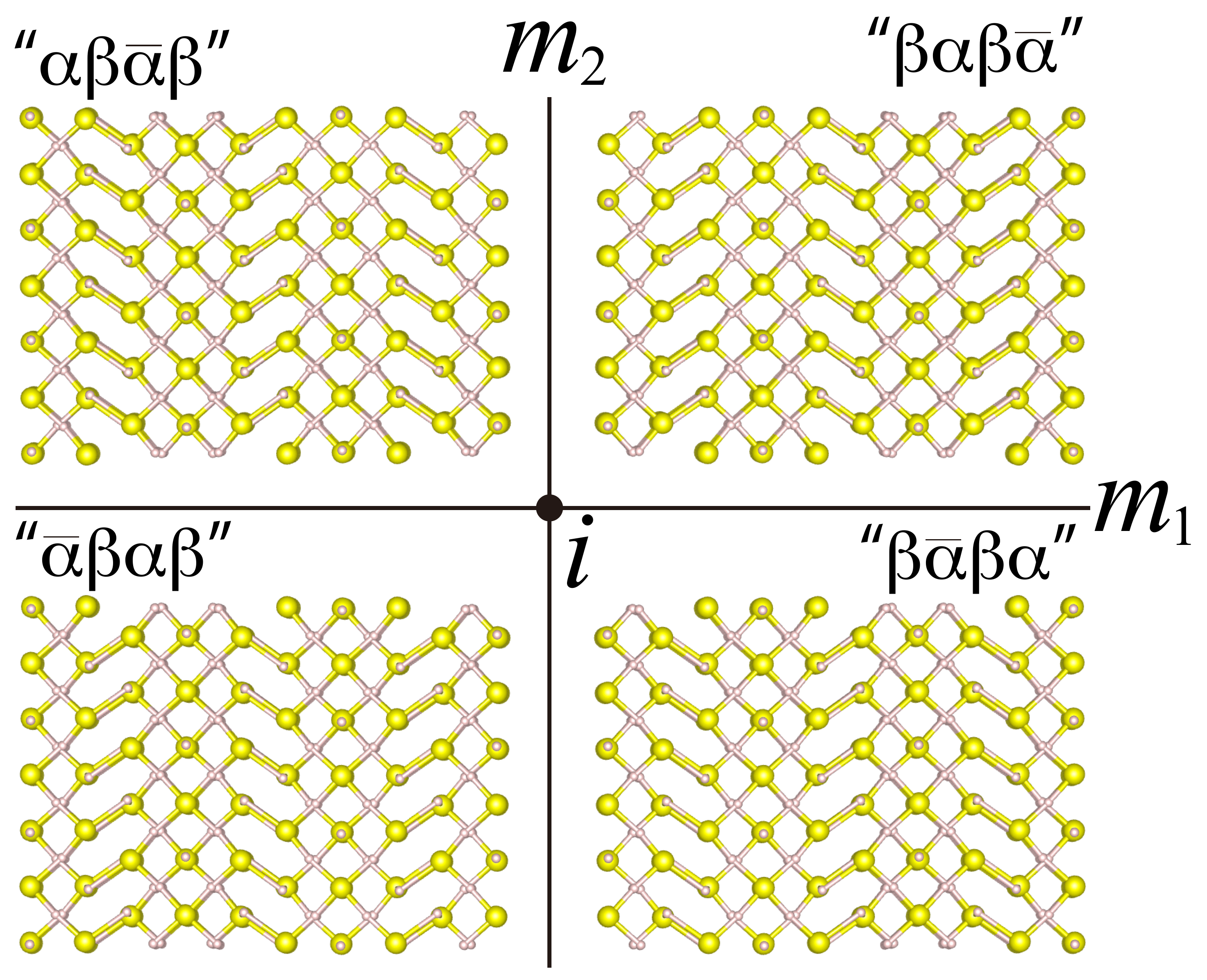}
  \caption{Relations between the operations on the permutation and structure, exemplified with permutation ``$\alpha\beta\bar{\alpha}\beta$".}
  \label{fig:permutation-op}
 \end{center}
\end{figure}

\subsection{Equivalence between the bonding permutations}
\label{sec:rule}
Here we summarize which permutations yield equivalent and nonequivalent metastable structures. (i) the cyclic variants of any permutations give equivalent structures because of the periodicity of the crystal. (ii) Pairs of permutations exchanged by $\alpha \leftrightarrow \bar{\alpha}$ (e.g., ``$\alpha\bar{\alpha}\beta$" and ``$\bar{\alpha}\alpha\beta$") yield pairs of structures exchanged by the mirror plane $m_{1}$ (plane perpendicular to the units, see Fig.~S\ref{fig:permutation-op}). (iii) Pairs of permutations exchanged by inversion (e.g., ``$\alpha\bar{\alpha}\beta$" and ``$\beta\bar{\alpha}\alpha$") yields structures interchanged by global inversion ($i$). All the permutations related by these three operations yield equivalent structures in that they give equal values of formation enthalpy. Note that the mirror reflection with respect to plane $m_{2}$ (plane parallel to the units, see Fig.~S\ref{fig:permutation-op}) is represented by the combination of the former operations.

\subsection{Relaxed structures}
We carried out the first-principles structure optimization for the structures summarized in Table \ref{tab:struct}, which resulted in the values of the formation enthalpy depicted in Fig.~4 (triangle). Note that this list exhausts all the possible structures for $N\leq 4$, in that any unlisted permutations can be related to either of the listed ones with the operations summarized in Fig.~S\ref{fig:permutation-op}.

The first-principles structure optimization has also been done for other compounds for reference. The squares in Fig.~4 correspond to the body-centered monoclinic (without spatial modulation) and $\beta$-Po sulfur (Ref.~\onlinecite{Degtyareva-sulfur}), $I4_{1}/amd$- and $C2/m$-HS (Ref.~\onlinecite{Errea-anharm-PRL}), $Pnma$-H$_4$S$_3$ (Ref.~\onlinecite{Li-Errea-various-struct-arxiv}), $P\bar{1}$- and $Cmca$-H$_2$S (Ref.~\onlinecite{Li-Ma-H2S-struct}), $R3m$- and $Im\bar{3}m$-H$_3$S (Ref.~\onlinecite{Duan-H3S-struct}), and $C2/c$-H (Ref.~\onlinecite{Pickard-Needs-H2}).

\begin{table}[b]
\caption[t]
{The calculated structures in terms of the corresponding permutation and the composition of the unit cell.}
\begin{center}
\label{tab:struct}
\tabcolsep = 1mm
\begin{tabular}{|l |c|c|} \hline
 $N$& Permutation & Composition (/unit cell) \\ \hline
2 & $\alpha\beta$& H$_7$S$_3$ \\ \hline
3 & $\alpha \alpha \bar{\alpha}$& H$_{12}$S$_6$ \\ \hline
3 & $\alpha \alpha \beta, \alpha \bar{\alpha} \beta$& H$_{11}$S$_5$ \\ \hline
3 & $\alpha \beta \beta$& H$_{10}$S$_4$ \\ \hline
4 & $\alpha \alpha \alpha \bar{\alpha}, \alpha \alpha \bar{\alpha} \bar{\alpha} $& H$_{16}$S$_8$ \\ \hline
4 & $\alpha \alpha \alpha \beta, \alpha \alpha \bar{\alpha} \beta, \alpha \bar{\alpha} \alpha \beta $& H$_{15}$S$_7$ \\ \hline
4 & $\alpha \alpha \beta \beta, \alpha \bar{\alpha} \beta \beta, \alpha \beta \bar{\alpha}  \beta $& H$_{14}$S$_6$ \\ \hline
4 & $\alpha \beta \beta \beta $& H$_{13}$S$_5$ \\ \hline
\end{tabular}
\end{center}
\end{table}

\subsection{Computational details}
The first-principles calculations were done using plane-wave pseudopotential calculation codes {\sc QUANTUM ESPRESSO}~\cite{QE}. The generalized-gradient approximation of the Perdew-Burke-Ernzerhof type~\cite{GGAPBE} was used for the exchange-correlation potential. We used the pseudopotentials for S and H atoms implemented with the Troullier-Martin scheme~\cite{TM-FHI}. The Kohn-Sham (KS) wave functions and charge density were described by the plane waves with energy cutoff of 80~Ry and 320~Ry, respectively.

The structure and lattice optimizations were done for calculating the formation enthalpy (Fig.~4) and electronic density of states (DOS) (Fig.~S\ref{fig:DOS}). We then described the charge density using the ${\bf k}$ points on the 4$\times$10$\times$10, 3$\times$10$\times$10, and 2$\times$10$\times$10 meshes  for the structures corresponding to the permutations of $N=2$, $N=3$, and $N\geq4$ (see a later paragraph for the setup of the lattices), respectively, with the first-order Hermite-Gaussian function of width of 0.03~Ry~\cite{M-P}. The values of the formation enthalpy were plotted for the structures up to $N=4$ (see Table~\ref{tab:struct}). The electronic densities of states for $P\bar{1}$-H$_2$S and $R\bar{3}m$-H$_3$S were calculated with the tetrahedron method~\cite{Bloechl-tetra} using the 33$\times$33$\times$23 and 33$\times$33$\times$33 {\bf k}-point meshes in the standard lattice setup~\cite{Li-Ma-H2S-struct,Duan-H3S-struct}, respectively, whereas those for the intermediate structures corresponding to the permutations $N=2,3,4,8,16$ were based on the 8$\times$40$\times$40 (for $N=2,3,4$), 6$\times$40$\times$40 (for $N=8$), 4$\times$28$\times$28 (for $N=16$) meshes, respectively. The X-ray diffraction simulations were done using RIETAN-VENUS package~\cite{RIETAN} for the optimized structures. The wavelength was set to 0.41377~\AA.

The superconducting transition temperature (Fig.~1) was calculated based on the density functional theory for superconductors~\cite{SCDFTI,SCDFTII}. The phonon-mediated and plasmon-assisted superconducting mechanisms are considered from the first-principles. In this case, we described the charge density with the the ${\bf k}$ points on finer meshes, based on which the crystal structures were optimized. We solved the gap equation given by Eq.~(5) in Ref.~\onlinecite{SCDFTII}. We employed the $n{\bf k}$-averaged approximation for the phononic kernels with the dependence on the density of states retained~[Eq.~(23) in Ref.~\onlinecite{SCDFTII} and Eq.~(40) in Ref.~\onlinecite{Akashi-Z-asym}, respectively]. We used the $n{\bf k}$-resolved formulation within the random-phase approximation (RPA) for the electronic kernel as described in Ref.~\onlinecite{Akashi-HxS}, in which the plasmon effect~\cite{Takada-plasmon,Akashi-plasmon-PRL, Akashi-plasmon-JPSJ} is included. The self-consistent potential responses for the lattice displacements were calculated with the density functional perturbation theory~\cite{Baroni}, with which the phonon frequencies, electron-phonon coupling matrix elements and the Eliashberg function $\alpha^{2}F(\omega)$ were calculated. We calculated the RPA dielectric matrix~\cite{Pines-RPA} with a reduced plane-wave energy cutoff of 12.8~Ry. The Brillouin-zone integrations involving the theta and delta functions in the phonon and dielectric-matrix calculations were done with the tetrahedron methods. The $n{\bf k}$ points for solving the SCDFT gap equation were generated with the sampling scheme described in Ref.~\onlinecite{Akashi-MNCl}, for which the KS energy eigenvalues and the electronic kernel were obtained by linear interpolation from the data on uniform grids. Further details are summarized in Table~\ref{tab:conditions}.

\begin{table*}[t!]
\caption[t]
{Detailed settings for the calculation of $T_{\rm c}$ in the $\alpha\beta\beta\beta$ structure. Subscript ``1" for {\bf q} points denotes the mesh with displacement by half a grid step.}
\begin{center}
\label{tab:conditions}
\tabcolsep = 1mm
\begin{tabular}{|l |c|c|c|c|c|} \hline
 &&$\alpha\beta$&$\alpha\beta\beta$&$\alpha\beta\beta\beta$&$\alpha(\beta\times9)$  \\ \hline
charge density &{\bf k} & (6 14 14) &(4 10 10)&(4 10 10)& (4 10 10)\\ \cline{2-6}
                      &interpol. &\multicolumn{4}{|c|}{1$^{\rm st}$ order Hermite-Gaussian~\cite{M-P} with width=0.030Ry} \\ \hline
dynamical matrix &{\bf k} & (6 14 14)&(4 10 10)&(4 10 10)&(4 10 10) \\ \cline{2-6}
                        &{\bf q} & (3 7 7)$_{1}$ &(2 5 5)$_{1}$&(2 5 5)$_{1}$&(2 5 5)$_{1}$ \\ \cline{2-6}
                        &interpol. &\multicolumn{4}{|c|}{Optimized tetrahedron~\cite{opt-tetra}} \\ \hline
electron-phonon &{\bf k}$^{\dagger}$ & (12 28 28) &(8 20 20)&(8 20 20)& (8 20 20)\\ \cline{2-6}
                        &interpol. &\multicolumn{4}{|c|}{Optimized tetrahedron~\cite{opt-tetra}} \\ \hline
dielectric function &{\bf k}  for bands crossing $E_{\rm F}^{\dagger\dagger}$ & (7 21 21) &(6 18 18)$_{1}$&(6 18 18)$_{1}$&(6 18 18)$_{1}$ \\ \cline{2-6}
                           &{\bf k}  for other bands& (3 7 7) &(2 6 6)$_{1}$&(2 6 6)$_{1}$&(2 6 6)$_{1}$ \\ \cline{2-6}
                           &{\bf q}  &  (3 7 7) &(2 6 6)&(2 6 6)&(2 6 6) \\ \cline{2-6}
                           &unoccupied band num.& 83 &105&134&301\\ \cline{2-6}
                           &interpol. &\multicolumn{4}{|c|}{Tetrahedron with the Rath-Freeman treatment~\cite{Rath-Freeman}} \\ \hline
DOS for phononic kernels &{\bf k}   & (17 27 27)  &(13 27 27)&(9 27 27)&(5 27 27)\\ \cline{2-6}
                           &interpol. &\multicolumn{4}{|c|}{Tetrahedron with the Bl\"{o}chl correction~\cite{Bloechl-tetra}} \\ \hline
SCDFT gap function & unoccupied band num. &39 &50&54&121 \\ \cline{2-6} 
           &{\bf k} for the electronic kernel & (3 7 7) &(2 6 6)$_{1}$&(2 6 6)$_{1}$&(2 6 6)$_{1}$ \\ \cline{2-6} 
           &{\bf k} for the KS energy eigenvalues & (17 27 27) &(13 27 27)&(9 27 27)&(5 27 27) \\ \cline{2-6} 
           & $N_{\rm s}$ for bands crossing $E_{\rm F}$ &6000 &5400&4500&2000 \\ \cline{2-6} 
           & $N_{\rm s}$ for other bands &200 &180&150&300,50,25\\ \cline{2-6}
           & Sampling error in $T_{\rm c}$ &\multicolumn{4}{|c|}{$\lesssim$3\%} \\ \hline 
           \multicolumn{6}{l}{$^\dagger$ Electron energy eigenvalues and eigenfunctions were calculated on these auxiliary grid points.} \\
           \multicolumn{6}{l}{$^{\dagger\dagger}$ Electron energy eigenvalues were calculated on these auxiliary grid points.} \\
\end{tabular}
\end{center}
\end{table*}

\begin{figure}[b!]
 \begin{center}
  \includegraphics[scale=0.25]{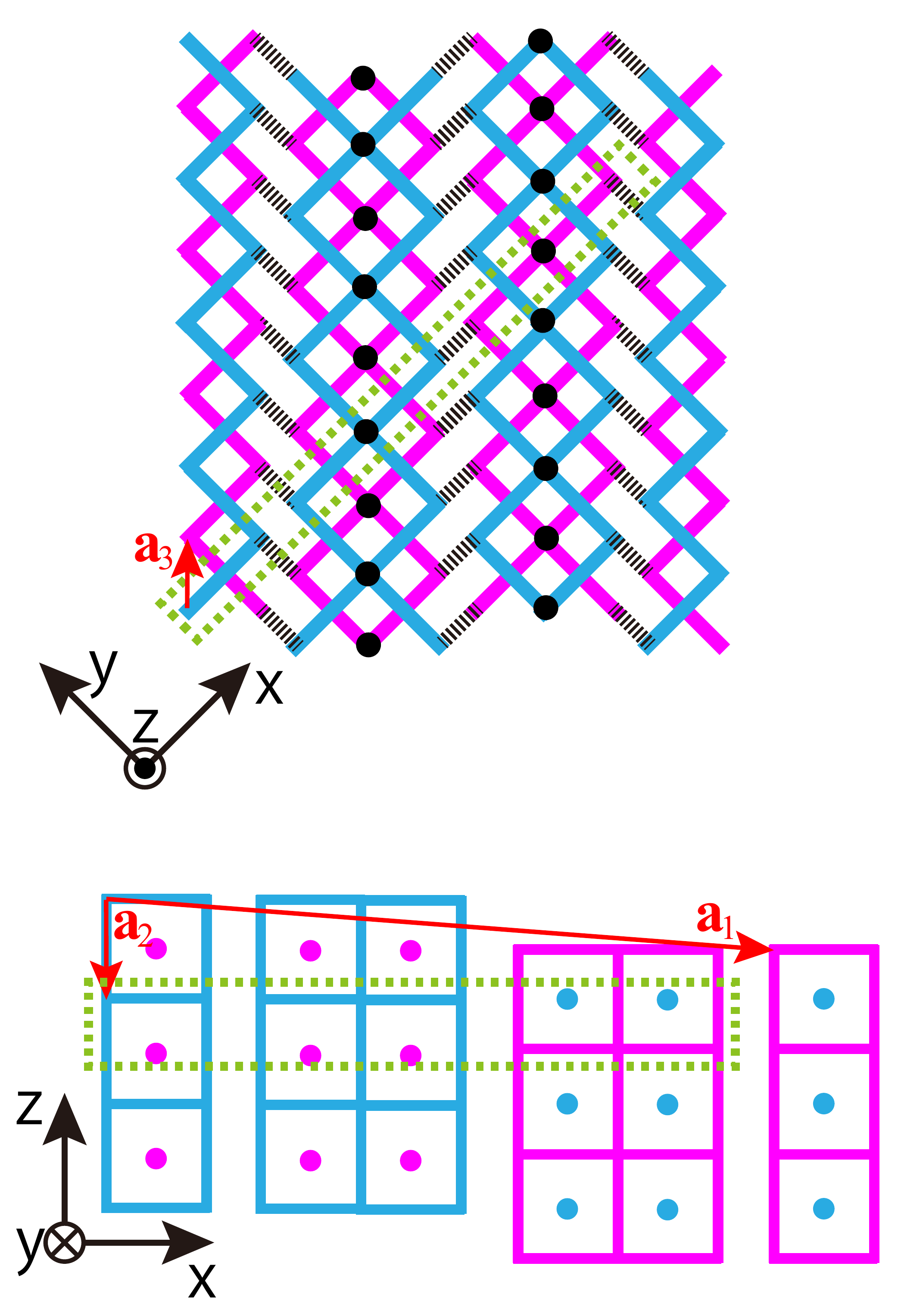}
  \caption{A scheme of implementing the initial structures, exemplified by the structure ``$\alpha\beta\bar{\alpha}\beta$". The rectangular area drawn with dashed line represents the unit cell, whereas the solid arrows represent the primitive lattice vectors defined in the present scheme. Note that the corners (centers of the edges) of the solid graph are occupied by sulfur (hydrogen) atoms.}
  \label{fig:lattice-implement}
 \end{center}
\end{figure}

We generated the input unit cells based on the following algorithm. (i) Define the S-H bonding length $a$, (ii) assuming the first element of the permutation to be $\alpha$, give the positions for two sulfur and four hydrogen atoms in the Cartesian coordinates by S$_{(1)}$=(0, 0, 0), S$_{(2)}$=(2, 0, 0), H$_{(1)}$=(0, 0, 1), H$_{(2)}$=(1, 0, 1), H$_{(3)}$=(2, 0, 1), H$_{(4)}$=(1, 0, 0) in unit of $a$. Here and hereafter, $A_{(n)}$ denotes the position of the $n$th atom of kind $A$. (iii) if the next element of the permutation is $\alpha$, generate two sulfur and four hydrogen atoms by S$_{(n)}$=S$_{(n-2)}$+(3, 0, 0) and H$_{(n)}$=H$_{(n-4)}$+(3, 0, 0) ; if $\bar{\alpha}$, generate two sulfur and four hydrogen atoms by S$_{(n)}$=S$_{(n-2)}$+(3, 0, 1$\times$$(-1)^{i}$) and H$_{(n)}$=H$_{(n-4)}$+(3, 0, 1$\times$$(-1)^{i}$) for the $i$th $\bar{\alpha}$; if $\beta$, on the other hand, generate one sulfur and three hydrogen atoms by S$_{(n)}$=S$_{(n-1)}$+(2, 0, 0) and H$_{(n)}$=H$_{(n-3)}$+(2, 0, 0). (iv) Repeat (ii)--(iii) to the end of the permutation, which yields the unit cell indicated by the dashed line in Fig.~S\ref{fig:lattice-implement}. (v) Finally define the three lattice vectors as ${\bf a}_{1}=(3N-k,0,  -l {\rm mod}2)$, ${\bf a}_{2}=(0, 0, -2)$, and ${\bf a}_{3}=(1, 1, 1)$ (see Fig.~S\ref{fig:lattice-implement}) for the permutation of period $N(\geq2)$ including $l$ $\bar{\alpha}$s and $k$ $\beta$s. We set $a=2.8$~Bohr for all the initial structures. We assumed no symmetry during the structure optimization, nor have we specified the symmetries of the final structures. Note that one can also develop different lattice-generation schemes yielding the same structures.

\subsection{Superconducting properties}
In Table~\ref{tab:el-ph-Tc}, we present the parameters representing the property of electron-phonon interaction, $\lambda$ and $\omega_{\rm ln}$, which are calculated from the Eliashberg function $\alpha^{2}F(\omega)$ with the formula $\lambda=2\int d\omega \frac{\alpha^{2}F(\omega)}{\omega}$ and $\omega_{\rm ln}={\rm exp}[\frac{2}{\lambda}\int d\omega \frac{\alpha^{2}F(\omega){\rm ln}\omega}{\omega}]$. Also, the resulting values of superconducting transition temperature by SCDFT $T_{\rm c}^{\rm SCDFT}$, which are also mentioned in the main text, are given. The renormalized Coulomb parameter $\mu^{\ast}$ was estimated from the Improved McMillan-Allen-Dynes equation~\cite{MAD} so that $T_{\rm c}^{\rm SCDFT}$ is reproduced using the calculated $\lambda$ and $\omega_{\rm ln}$.

\begin{table}[h]
\caption[t]
{First-principles calculated electron-phonon coupling parameters and superconducting transition temperature.}
\begin{center}
\label{tab:el-ph-Tc}
\tabcolsep = 1mm
\begin{tabular}{|l |c|c|c|c|} \hline
 &$\alpha\beta$&$\alpha\beta\beta$&$\alpha\beta\beta\beta$&$\alpha(\beta\times9)$  \\ \hline
 $\lambda$&0.936&1.162&1.303&1.478 \\ \hline
  $\omega_{\rm ln}$ (K)&893&1218&1289&1355 \\ \hline
   $T_{\rm c}^{\rm SCDFT} (K)$&41&80&107&121 \\ \hline
   $\mu^{\ast}$&0.160&0.173&0.164&0.191 \\ \hline
\end{tabular}
\end{center}
\end{table}

\subsection{Electronic density of states}
\begin{figure}[t!]
 \begin{center}
  \includegraphics[scale=0.5]{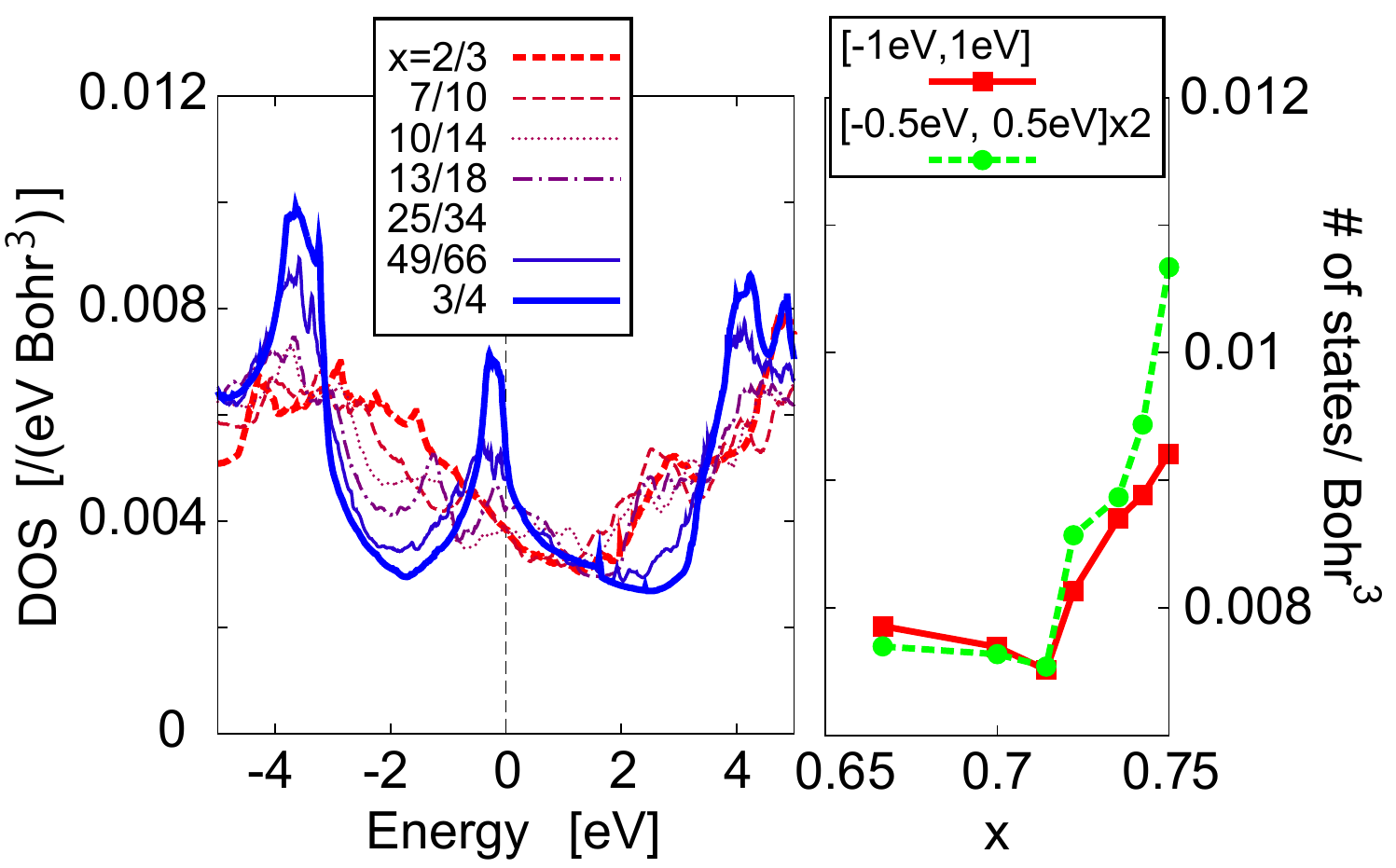}
  \caption{(left) Density of states (DOS) per unit volume for various compositions. (right) Total number of states per unit volume electron calculated within the range of [-1eV, 1eV] and [0.5eV, 0.5eV].}
  \label{fig:DOS}
 \end{center}
\end{figure}

The calculated structures of the DOS for the intermediate compositions H$_{x}$S$_{1-x}$ are shown in Fig.~S\ref{fig:DOS}. The finite values at the Fermi energy (vertical dashed line) indicates that all the systems are consistently metallic. We also observed a peak in the DOS near the Fermi level growing gradually with the increase of $x$, which evolves into the van-Hove peak found in the pristine $Im\bar{3}m$-H$_3$S\cite{Duan-H3S-struct, Mazin-enthalpy, Pickett-atomic}. More quantitatively, we also calculated the total number of states per valence electron within the range of $\pm$ 1~eV ($\simeq 5\times \omega_{\rm D}$) and $\pm$ 0.5~eV ($\simeq 2.5\times \omega_{\rm D}$) with $\omega_{\rm D}$ being the Debye frequency~\cite{Duan-H3S-struct}, respectively (Fig.~S\ref{fig:DOS}, right). These results indicate that the electronic structure is not linearly but at least continuously dependent on the H ratio $x$.

\subsection{On the possibility of polaron formation}
In the TiO$_{x}$ Magn\'{e}li phases at low temperature, static ``bipolaron" phase has been observed~\cite{Lakkis-bipolaron}, where small fraction of Ti atoms trap excess valence electrons and form local dimers. Proximity effects of this phase also affects the electronic transport even in the disordered phase at high temperature. The multivalence nature of Ti atoms obviously has a crucial role in this phenomenon. In the present H$_{x}$S case, the possible counterpart should involve hydrogen atoms for their multivalency.

We then carried out the Bader-charge analysis for the structures corresponding to the permutations up to period $N=4$. The calculations were performed with ``bader" code~\cite{Bader}. The result for those of the hydrogen atoms, which can exhibit multiple valence states H$^{\pm1}$ and H$^{0}$, is summarized in Table~\ref{tab:bader}. To show the charge distribution among them, we present two representative values:
\begin{eqnarray}
\bar{\rho}_{\rm H}
\equiv
\frac{1}{N_{\rm H}}
\sum_{i}^{N_{\rm H}} \rho_{{\rm H}_{(i)}}
,
\\
\Delta\rho_{\rm H}^{\rm max}
\equiv
{\rm max}_{ij}|\rho_{{\rm H}_{(i)}}-\rho_{{\rm H}_{(j)}}|
.
\end{eqnarray}
$N_{\rm H}$ denotes the total number of hydrogen atoms in the unit cell and $\rho_{H_{(i)}}$ is the value of the Bader charge at the $i$th hydrogen atom. $i$ and $j$ are indices of the hydrogen atoms in the unit cell. From the former quantity, we find that the hydrogen atoms are well regarded to be H$^{0}$, whereas the latter indicates that the charge imbalance among the hydrogen atoms are at maximum $\sim 0.15$. The tendency to form polarons is thus thought to be not so strong as in the titanium oxide~\cite{Lakkis-bipolaron}.
\begin{table}[h]
\caption[t]
{Mean values and maximum differences of the Bader charges for hydrogen atoms. For each atom, we attributed the value to that located at the nearest site.}
\begin{center}
\label{tab:bader}
\tabcolsep = 1mm
\begin{tabular}{|c|c|c|c|c|c|c|c|} \hline
$N$&bonding& \multicolumn{2}{|c|}{110GPa}&\multicolumn{2}{|c|}{150GPa}&\multicolumn{2}{|c|}{190GPa} \\ \hline
   &                     &$\bar{\rho}_{\rm H}$&$\Delta\rho_{\rm H}^{\rm max}$ &$\bar{\rho}_{\rm H}$&$\Delta\rho_{\rm H}^{\rm max}$ &$\bar{\rho}_{\rm H}$&$\Delta\rho_{\rm H}^{\rm max}$ \\ \hline
2 &$\alpha\beta$&0.9358&0.0692&0.9357&0.0387&0.9481&0.0299 \\\hline
3 &$\alpha\alpha\bar{\alpha}$&0.9234&0.0916&0.9226&0.0757&0.9446&0.0856 \\\hline
3 &$\alpha\alpha\beta$&0.9358&0.1240&0.9350&0.0829&0.9429&0.0949 \\\hline
3 &$\alpha\bar{\alpha}\beta$&0.9352&0.0814&0.9405&0.0316&0.9600&0.0514 \\\hline
3 &$\alpha\beta\beta$&0.9424&0.0732&0.9360&0.0503&0.9549&0.0365 \\\hline
4 &$\alpha\alpha\alpha\bar{\alpha}$&0.9187&0.1030&0.9224&0.0796&0.9419&0.0907 \\\hline
4 &$\alpha\alpha\bar{\alpha}\bar{\alpha}$&0.9199&0.1027&0.9238&0.1102&0.9483&0.1095 \\\hline
4 &$\alpha\alpha\alpha\beta$&0.9264&0.1520&0.9232&0.0643&0.9424&0.0657 \\\hline
4 &$\alpha\alpha\bar{\alpha}\beta$&0.9286&0.1216&0.9301&0.0724&0.9483&0.0721 \\\hline
4 &$\alpha\bar{\alpha}\alpha\beta$&0.9282&0.0841&0.9327&0.0359&0.9538&0.0336 \\\hline
4 &$\alpha\alpha\beta\beta$&0.9343&0.1028&0.9345&0.0966&0.9451&0.0547 \\\hline
4 &$\alpha\bar{\alpha}\beta\beta$&0.9388&0.0850&0.9428&0.0563&0.9583&0.0421 \\\hline
4 &$\alpha\beta\bar{\alpha}\beta$&0.9379&0.0641&0.9452&0.0427&0.9605&0.0260 \\\hline
4 &$\alpha\beta\beta\beta$&0.9508&0.0831&0.9452&0.0607&0.9549&0.0542 \\\hline
\end{tabular}
\end{center}
\end{table}


\subsection{Possible local reactions forming the H$_3$S slabs}
\begin{figure}[t!]
 \begin{center}
  \includegraphics[scale=0.35]{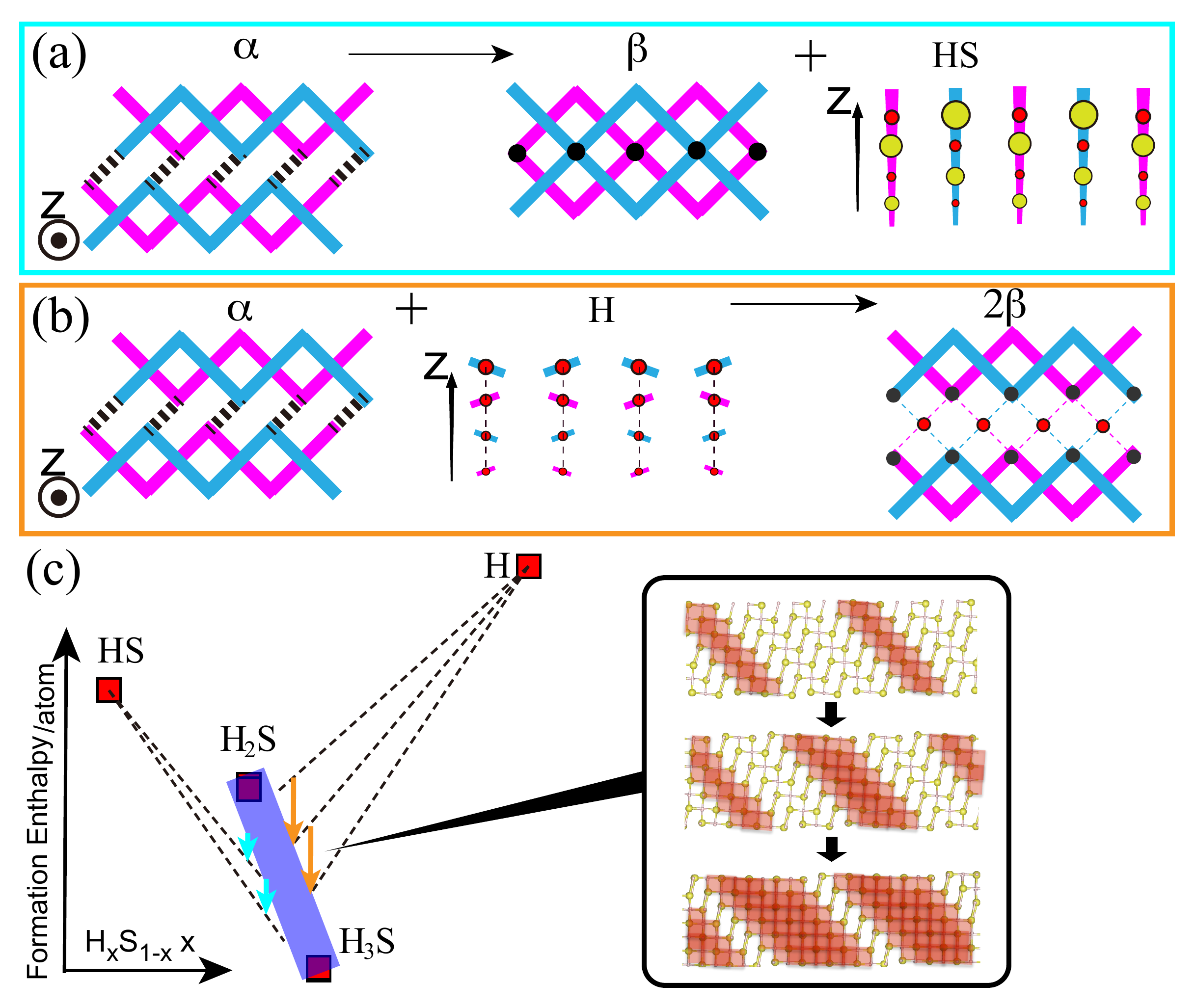}
  \caption{(a) Reaction $\alpha$ ($\bar{\alpha}$)$\rightarrow$$\beta$+HS and (b) Reaction $\alpha$ ($\bar{\alpha}$)+H$_{2}$$\rightarrow$2$\beta$, where small and large balls represent hydrogen and sulfur atoms, respectively. (c) Schematic picture of the possible transformation paths (see also Fig.~5).}
  \label{fig:transforms-summary}
 \end{center}
\end{figure}

Here we summarize the reaction(s) that can occur in the compressed H$_2$S and intermediate phases. As discussed in the main text, the pressure-induced stabilization of the HS and ``Magn\'{e}li" phases should stimulate a reaction $\alpha$ ($\bar{\alpha}$)$\rightarrow$$\beta$+HS [Fig.~S\ref{fig:transforms-summary}(a)]. There is also another possible reaction forming H$_3$S slabs $\alpha$ ($\bar{\alpha}$)+H$_{2}$$\rightarrow$2$\beta$ [Fig.~S\ref{fig:transforms-summary}(b)]. The values of formation enthalpy suggests that this reaction is also stimulated upon compression [Figs.~4 and S\ref{fig:transforms-summary}(c)]. Although the latter reaction requires excess hydrogen atoms, it may also have been relevant in the previous experiments: Observation of phases of elemental sulfur~\cite{Eremets,Shimizu-struct-arxiv} indicates H$_2$ or any other H-rich phases as their decomposition counterparts.

\begin{thebibliography}{999}
\bibitem{comment-distortion} Note that the theoretically predicted $R3m$ distortion is so small as to be difficult to measure in the experiment.
\bibitem{comment-bonding} We can also regard that the dominant inter-unit bonding is the S-S one~\cite{Li-Ma-H2S-struct} instead S-H, though this difference does not change the conclusion of the later discussions.
\bibitem{comment-reaction} We can also consider another possible reaction forming local H$_3$S slabs and stimulated by compression: $\alpha$ ($\bar{\alpha}$)+H$_{2}$$\rightarrow$2$\beta$~\cite{comment-suppl}.
\end{thebibliography}
\end{document}